\documentclass[hyper,12pt,letterpaper]{JHEP3}
\usepackage{amsmath,amssymb,multirow,array}

\numberwithin{equation}{section}

\newcommand{\nn}{\nonumber}

\title{Anomalous charged fluids in 1+1d from equilibrium partition function}
\author{Sachin Jain$^a$, Tarun Sharma$^a$\\
$^a$Dept. of Theoretical Physics, Tata Institute of Fundamental Research, Homi Bhabha Rd,
Mumbai 400005, India. \\
Email:\ \ {\bf sachin@theory.tifr.res.in, tarun@theory.tifr.res.in
} }

\abstract{In this note we explore the constraints imposed by the existence of equilibrium partition on  parity violating 
charged fluids in 1+1 dimensions at zero derivative order. We write the  
equilibrium partition function consistent with 
1+1 dimensional CPT invariance and which reproduces the correct anomaly in the charge current. The constraints on constitutive
relations obtained in this way 
matches precisely with those obtained using the second law of thermodynamics.}

\begin{document}
\maketitle

\section{Introduction}
Relativistic Fluid dynamics is an effective large wavelength description (at length scales much bigger than the 
mean free path) of certain phases of matter which at microscopic level are described by relativistic quantum field theories. 
The basic equations governing dynamics in this description are the conservation laws corresponding to the global symmetries of the 
underlying theory. More specifically, these are the conservation equations of the stress energy tensor and charge currents. 
 These equations are to be supplemented by constitutive relations which expresses the stress-energy tensor and charge current 
in terms of the basic fluid variables namely velocity, temperature  and chemical potential. 

Consistency with second law of thermodynamics has been used as a constraining principle on the constitutive relations in 
fluid dynamics (see \cite{Landaufd,sonsurowka,Bhattacharya:2011tra,Bhattacharya:2011eea,Bhattacharyya:2012ex,Loganayagamanom,Loganayagam:2012pz} 
and references therein). This gives two kinds of relations: a) Inequality 
type relations on dissipative coefficients(which contributes to entropy increase). b) Equality type relations on non dissipative 
coefficients(which do not contribute to entropy increase). Recently in \cite{Banerjeefdpf}(see also \cite{Jensen:2012jh}) it was shown that the requirement of the existence of 
stationary equilibrium which is generated from a partition function gives all the equality type relations. One of the cases that was 
studied in \cite{Banerjeefdpf}  was charged fluid dynamics in 3+1 dimensions when the charge current is anomalous.
 In this case, the results of Son and Surowka  \cite{sonsurowka}  on
the chiral magnetic and chiral vorticity flows, were recovered without making any reference to
an entropy current.

%

 In this note we study the anomalous charged fluid dynamics in 1+1 dimensions   using the equilibrium partition function. 
This system has earlier been studied in \cite{dubovsky2d} using the second law of thermodynamics as well as
 from an action point of view. In this note we write down the equilibrium partition function for this system at zero derivative 
order which reproduces the anomalous charge conservation and on comparison with the  most general constitutive relations in fluid 
dynamics, gives the results obtained in \cite{dubovsky2d}.

\section{1+1d parity violating charged fluid dynamics} \label{pf}
Consider the parity violating charged fluids in 1+1 dimensions with background metric and gauge field 
\begin{equation}\label{bg}\begin{split}
ds^2 &= -e^{2\sigma} (dt + a_1 dx)^2 + g_{11} dx^2 \\
{\cal A} &= {\cal A}_0 dt + {\cal A}_1 dx^1  ~~.             
\end{split}
\end{equation}

The equations of motion are the following anomalous conservation laws
\begin{equation}\begin{split}\label{eom}
\nabla_{\mu} T^{\mu\nu} = {\cal F}^{\nu\lambda} \tilde{J}_{\lambda} \\
\nabla_{\mu} \tilde{J}^{\mu} = c \epsilon^{\mu\nu} {\cal F}_{\mu\nu} \\
\nabla_{\mu} J^{\mu} = \frac{c}{2} \epsilon^{\mu\nu} {\cal F}_{\mu\nu} \\
\end{split}
\end{equation}
here  $\tilde{J},~ J$ are covariant and consistent currents respectively (\cite{Bardeenanom}, see also \cite{Banerjeefdpf}).

The most general partition function consistent with Kaluza-Klein gauge invariance\footnote{\begin{equation}
V'_i=
V_i- \partial_i \phi V_0,
\quad (V')^0=
V^0+ \partial_i \phi V^i.
\end{equation}}, diffeomorphism 
along the spatial direction and $U(1)$ gauge invariance upto anomaly is 
\begin{equation}\label{pfgen}\begin{split}
{\cal W} &= {\cal W}_{inv} + {\cal W}_{anom} \\
{\cal W}_{inv} &= C_1 T_0 \int A_1 dx - C_2 T_0 \int a_1 dx \\
{\cal W}_{anom} &= -\frac{C}{T_0} \int A_0 A_1 dx 
\end{split}
\end{equation}
where $C$, $C_1$ and $C_2$ are constants independent of $\sigma$ and $A_0$  and  
\begin{equation}
 A_0 = {\cal A}_0 + \mu_0,~~A_i ={\cal A}_i- A_0 a_i.
\end{equation}
Equation \ref{pfgen} is written in terms of $A_i$ which unlike ${\cal A}_i$, are Kaluza-Klein gauge invariant.

Under $U(1)$ gauge transformation $ A_0 \rightarrow A_0,~ A_1 \rightarrow A_1 + \partial_1 \phi,$  
we obtain\footnote{Since we are interested in time independent background fields, we consider only time independent gauge transformations.} 
\begin{equation}\label{wgt}\begin{split}
\delta {\cal W}_{inv} &= 0 \\
 \delta {\cal W}_{anom} &= \frac{C}{T_0} \int \phi~\partial_1 A_0 dx = -\frac{C}{2} \int d^2x \sqrt{-g_2}~\phi~\epsilon^{\mu\nu} {\cal F}_{\mu\nu}.
\end{split}
\end{equation}

Table (\ref{cpttab}) lists the action of 2 dimensional C, P and T on various fields. 
 Requiring CPT invariance  sets 
$C_1$ to zero since the term with coefficient $C_1$ is odd under CPT.
\begin{table}
\centering
\begin{tabular}[h]{|c|c|c|c|c|}
\hline
Field & C & P & T & CPT \\
\hline
$\sigma$ & $+$ & $+$ & $+$ & $+$ \\
\hline
$a_1$ & $+$ & $-$ & $-$ & $+$ \\
\hline
$g_{11}$ & $+$ & $+$ & $+$ & $+$ \\
\hline
$A_0$ & $-$ & $+$ & $+$ & $-$ \\
\hline
$A_1$ & $-$ & $-$ & $-$ & $-$ \\
\hline
\end{tabular}
\caption{Action of CPT}
\label{cpttab}
\end{table}

Now let us look at the most general  constitutive relations  allowed by symmetry in the parity violating 
case at zero derivative order. At this order, there are no gauge invariant parity odd scalar 
or tensor.  But one can construct a gauge invariant vector \footnote{In components the parity odd vector is 
\begin{equation}\label{uteq}
{\tilde u}_0 =0,~~{\tilde u}^1=\epsilon^{10}u_{0} =\epsilon^{1} 
\end{equation}
where $\epsilon^{1} =e^{\sigma} \epsilon^{01} = \frac{1}{\sqrt{g_{11}}}.$
}
\begin{equation}\label{udual}
{\tilde u}^{\mu}=\epsilon^{\mu\nu}u_{\nu}.
\end{equation}

The most general  allowed constitutive relations allowed by symmetry in Landau frame thus take the form
\begin{equation} \label{cr}\begin{split}
T^{\mu\nu} &= (\epsilon + p) u^\mu u^\nu + p g^{\mu\nu} \\
\tilde{J}^\mu &= q u^{\mu} + \xi_j \tilde{u}^{\mu}. \\
\end{split}
\end{equation}


\subsection{Equilibrium from  Partition Function}\label{2deqpf}
In this subsection we will use the equilibrium partition function \eqref{pfgen} to obtain 
the stress tensor and charge current at zero derivative order. 
Setting $C_1$ to zero in \eqref{pfgen} we have 
\begin{equation}\label{pf}
{\cal W} = -\frac{C}{T_0} \int A_0 A_1 dx- C_2 T_0 \int a_1 dx
\end{equation}

Using the partition function \eqref{pf} it is straightforward to compute the stress tensor and 
charge current\footnote{\begin{eqnarray}
T_{00}&=& -\frac{T_0 e^{2 \sigma}}{\sqrt{-g_{(p+1)}} }\frac{\delta W}{\delta \sigma},
\quad T_0^i= \frac{T_0}{\sqrt{-g_{(p+1)}} }\bigg(\frac{\delta W}{\delta a_i}-
A_0 \frac{\delta W}{\delta A_i}\bigg) , \nn \\
T^{ij}&=& -\frac{2 T_0}{\sqrt {-g_{(p+1)}}} g^{il}g^{jm}\frac{\delta W}
{\delta g^{lm}}, \quad
J_0 =-\frac{e^{2\sigma} T_0}{\sqrt{-g_{(p+1)}}}\frac{\delta W}{\delta A_0},
\quad J^i=\frac{T_0}{\sqrt{-g_{(p+1)}}}\frac{\delta W}{\delta A_i}.
\end{eqnarray}
where, for instance, the derivative w.r.t $A_0$ is taken at constant $\sigma$, $a_i$, $A_i$,
$g^{ij}$, $T_0$ and $\mu_0$. See  \cite{Banerjeefdpf} for details.} in equilibrium to be 
\begin{equation}\begin{split}\label{eqstcr}
T_{00} &= 0,~~ T^{11} = 0, ~~
T_0^1 = e^{-\sigma}\epsilon^1 \left(- T_0^2 C_2 + C A_0^2 \right), \\
J_0 &= C \epsilon^1 A_1 e^{\sigma},~~
J^1 = -C \epsilon^1 e^{-\sigma} A_0. 
\end{split}
\end{equation}
The covariant current ($\tilde{J}^\mu$) can be obtained from the consistent current ($J^\mu$) by 
an appropriate shift as follows
\begin{equation}
\tilde{J}^\mu = J^{\mu} + J^{\mu}_{sh}, ~~~ J^{\mu}_{sh} = C \epsilon^{\mu\nu} A_{\nu}.
\end{equation}
In components the covariant current is then 
\begin{equation}\label{cocur}
\tilde{J}_0 = 0, ~~ \tilde{J}^1 = -2C e^{-\sigma}\epsilon^{1} A_0 .
\end{equation}

\subsection{Equilibrium from Hydrodynamics}

\begin{table}
\centering
\begin{tabular}[h]{|c|c|c|}
\hline
Type & Data& Evaluated at equilibrium \\
 & & $T=T_0e^{-\sigma}, ~ \mu=e^{-\sigma} A_0, ~ u^\mu= u^\mu_K$ \\
\hline
 Scalars & None  & None \\
\hline
 Vectors & $u^{\mu}$  & $\delta^{\mu}_{0}e^{-\sigma}$ \\     
\hline
Pseudo-Vectors & $\epsilon_{\mu\nu}u^{\nu}$ & $ \epsilon_{1}$ \\
\hline
Tensors & None &None \\
\hline
\end{tabular}
\caption{Zero derivative fluid data}
\label{zdfd}
\end{table}

\begin{table}
\centering
\begin{tabular}[h]{|c|c|}
\hline
Scalars & None  \\
\hline
Vectors & none ,  none \\
\hline
Pseudo-Vectors & $\epsilon^{1} f(\sigma,A_0)$ \\
\hline
Tensors & None \\
\hline
\end{tabular}
\caption{Zero derivative background data}
\label{zdbd}
\end{table}

We are interested in the stationary equilibrium solutions to conservation equations corresponding 
to the constitutive relations \eqref{cr}.
 The equilibrium solution in the parity even sector in background \eqref{bg} at zero derivative order is 
 \begin{equation}\label{pesol}
 u^{\mu} = u^{\mu}_{(0)} = e^{-\sigma}(1,0),~~ T = T_0 e^{-\sigma}, ~~ \mu = A_0 e^{-\sigma}.
 \end{equation} Since there are no gauge invariant parity odd scalars 
in table \ref{zdbd}, temperature and chemical potential do not receive any correction. However, the fluid 
velocity in equilibrium receives correction as 
\begin{equation}\label{ucor}
u^{\mu} = u^{\mu}_{(0)} + b \epsilon^{\mu\nu} u^{(0)}_{\nu}.
\end{equation}

From \eqref{cr}, \eqref{pesol} and \eqref{ucor} we get the parity odd correction to the equilibrium stress tensor and 
charge current, which receive contribution from correction to the constitutive relations as well as from correction to the
equilibrium fluid velocity, to be 
\begin{equation}\label{ccvm} \begin{split}
\delta T_{00}&=\delta J_0=\delta T^{ij}=0,\\
\delta T_{0}^1 &=-e^{\sigma}(\epsilon+P)b \epsilon^1, \\
\delta {\tilde J}^1 &= (q b + \xi_j) \epsilon^1 .\\
\end{split}
\end{equation}

\subsection{Constraints on Hydrodynamics}
Comparing the non trivial components of the equilibrium stress tensor and charge current of \eqref{eqstcr} 
and \eqref{ccvm} we find that the coefficient of velocity correction \eqref{ucor} is 
\begin{equation}\label{vcor}
b = -\frac{T^2}{\epsilon+ p}\left(-C_2 + C\nu^2\right) \\
\end{equation}
and the coefficient in correction to charge current \eqref{cr} is 
\begin{equation}\label{SSsolv}
\xi_{j} = C\left(\frac{q \mu^2}{\epsilon+p}-2\mu\right)-C_2 \frac{q T^2}{\epsilon+p}.
\end{equation}

where $\nu = \frac{\mu}{T}=\frac{A_{0}}{T_{0}}$.

The expressions \eqref{SSsolv} agree exactly with
the results of  \cite{dubovsky2d} based on the requirement of positivity of the
entropy current and effective action.

\subsection{The Entropy Current}
The equilibrium entropy can be obtained from the partition function using
\begin{equation}\label{fent}
\begin{split}
S&=\frac{\partial}{\partial T_0}(T_0 \log Z) \\
&=-2 C_2 T_0\int  \sqrt{g_{11}}\epsilon^{1}a_1 dx~.
\end{split}
\end{equation}

In this subsection we determine the constraints on the hydrodynamical entropy current $J^\mu_S$ 
from the requirement that \eqref{fent} agree with the local integral
\begin{equation}\label{entn}
S=\int d x \sqrt{-g_2} J^0_S~~.
\end{equation}

The most general form of the entropy current  allowed by symmetry
\footnote{Let us note that the entropy current need not be gauge invariant, see \cite{Banerjeefdpf} for more details.},
 at zero derivative order is 
\begin{equation}\label{mjecn}\begin{split}
J_S^\mu=&~s u^\mu + \xi_{s} {\tilde u}^{\mu} + h \epsilon^{\mu\nu}{\cal A}_\nu ~,\\
\end{split}
\end{equation}
where $h$ is a constant.

 The parity odd correction to the entropy current in equilibrium, which receives 
contributions  both from correction to the hydrodynamical entropy current and equilibrium velocity, is given by
\begin{equation}\label{zerocompent}
J_S^0|_{correction} = s \delta u^0 + \xi_s {\tilde u}^0 + h \epsilon^{01}{\cal A}_1.
\end{equation}

Now using
$$ \nu =\frac{ A_0}{T_0}, ~~\delta u^0 =- a_1 \delta u^1=-b\epsilon^1a_1, ~~{\tilde u}^{0} =-\epsilon^1 a_1 $$
the  correction to the hydrodynamical entropy in equilibrium is given by 
\begin{equation}\label{totalent}
\int d x \sqrt{-g_2} J^0_s|_{correction} = \int d x~ e^{\sigma}\left((-s b-\xi_s)\epsilon^1 a_1 + h \epsilon^1(A_1 + A_0 a_1)\right).
\end{equation}

Comparing this expression with \eqref{fent} and using \eqref{SSsolv} we find
 \begin{equation}\label{naammil}
 \xi_s = C\frac{s \mu^2}{\epsilon+p}+ C_2 T \left( 1+\frac{\rho\mu}{\epsilon+p}\right) ,~~~~h = 0.
 \end{equation}
This result is in precise agreement with those of \cite{dubovsky2d}.

\section{Conclusion}
To conclude, for $1+1$d parity violating charged fluid in a time independent background with 
anomaly  one can write down a local equilibrium partition function and 
the constraints obtained on the constitutive relations by demanding consistency with this partition 
function are in agreement with those obtained from a local form of entropy increase principle. In \cite{Banerjeefdpf}, 
by demanding the existence of
a partition function it was noted that,
for first order $2+1$d parity violating charged fluid and second order $3+1$d uncharged fluid, one obtains weaker 
constraints on the non dissipative part of
the entropy current as compared to that obtained by demanding entropy increase. However, for the case of first order $3+1$d 
charged fluid with anomaly, the entropy current obtained in both ways agree so is also for $1+1$d anomalous case, as shown in this note. 
It would be interesting to check this for an anomalous fluid in arbitrary
dimensions.


\acknowledgments
We would like to thank Nabamita Banerjee for collaboration in some parts of this work. We would also 
like to thank Shiraz Minwalla, Sayantani Bhattacharyya, R. Loganayagam, V. Umesh for useful discussions and comments. 
We would also like to acknowledge our debt to the
people of India for their generous and steady support to research in the basic sciences.

\bibliographystyle{JHEP}
\bibliography{2dref}

\end{document}